# Inconsistency of the photoemission spectrum with the spectral function in Kondo systems

S. Patil[1], A. Generalov[1] & A. Omar[2]

[1] Institut für Festkörperphysik, Technische Universität, Dresden, D-01062 Dresden, Germany.
[2] Leibniz-Institut für Festkörper- und Werkstoffforschung (IFW) Dresden, D-01171 Dresden, Germany.

**Kondo effect**[1] **is an example of asymptotic freedom**[2,3] **in condensed matter where the strength of the antiferromagnetic coupling between a localized magnetic moment and valence electrons, increases with reducing temperature. The electronic spectral function for such a system predicts a narrow Kondo resonance**[4] **close to Fermi energy below its Kondo temperature '$T_K$', signalling the formation of quasiparticles. The internal energy level structure of the localized magnetic moment introduces sidebands near the Kondo resonance, each with its respective Kondo temperature increasing as the sideband position moves towards higher binding energies**[5]**. Consequently, the temperature dependence of the resonance features becomes weaker as we go towards higher binding energies, which is predicted by the spectral function**[6]**. Here, we show a temperature evolution of the Ce4$f$ photoemission spectrum obtained from a prototypical Kondo system CeAl$_2$, departing from the predictions of the spectral function. Our measurements reveal a uniform temperature dependence for all the well screened Ce4$f$ photoemission features (4$f^1$ final state), which are usually interpreted as resonance features, as we reduce the temperature suggesting that the resonance interpretation for the 4$f^1$ final state features is not appropriate. Instead, the spectral evolution is explained simply by the temperature dependence of the strength of the coupling between Ce4$f$ and valence electrons, which remains uniform over the range of 4$f^1$ final state features. As a possible explanation to the observed spectral evolution, we propose the phenomenon of collapse**[7] **of the Kondo singlet wave function upon photoelectron kinetic energy measurement, analogous to the wave function collapse observed upon measurements performed on an entangled EPR**[8] **(Einstein, Podolsky, Rosen) pair. Our proposal highlights how experiments studying single particle properties, when performed on a quantum system, give incomplete information about the many-body physics of the system.**

Understanding the photoemission spectra from rare earth based Kondo systems has been a long standing puzzle with conflicting opinions emerging over time. The question was 'how to understand the temperature evolution of the electronic structure in Kondo systems and how to see the signature of Kondo resonance in photoemission, a many-body feature representing the formation of Kondo singlet state', at low temperatures. Early reports about the observation of Kondo resonance in temperature dependent photoemission studies[9,10] were challenged and explained alternatively[11,12]. The question was whether one needs a many-body approach like the single impurity Anderson model[13] (SIAM; describes Kondo physics by means of hopping between 4$f$ and valence states) to explain the temperature evolution of Ce4$f$ photoemission features or one can explain it merely using the conventional thermal broadening mechanisms like phonon/Fermi broadening etc. With the advent of high resolution photoemission spectrometers, more high resolution photoemission studies were done, which favoured the SIAM picture for understanding the Ce4$f$ spectral evolution[5,14,15] and the question seem to have been answered in favour of SIAM. In earlier reports, the focus was mainly on studying the temperature evolution of the well screened Ce4$f$ photoemission features believed to be located within the binding energy region of 0.5 eV from the Fermi level. However, a recent report on the temperature dependent x-ray photoemission spectra from a prototypical Kondo system CeB$_6$, collected for a wider energy range extending beyond 2 eV binding energy, showed an unusual Ce4$f$ spectral evolution where the well screened Ce4$f$ photoemission features show an enhancement with respect to the poorly screened Ce4$f$ photoemission feature at low temperatures, if one normalizes the spectra at the peak intensity of the poorly screened feature[16]. Interestingly, this enhancement is seen to extend to higher binding energies beyond 0.5 eV from the Fermi level, up to the poorly screened Ce4$f$ photoemission feature, thus suggesting that the well screened Ce4$f$ photoemission features seem to extend to higher binding energies than 0.5 eV, contrary to the previous belief. Such an enhancement is not an artefact of the Fermi/phonon broadening of the photoemission features, as was raised in ref. 11 and arises due to the many-body physics of the system[16]. Therefore it is of interest to explore the finer details of such a high binding energy enhancement with high energy resolution and by exploiting the tunability of photon energies in synchrotron radiation sources to selectively study the unusual Ce4$f$ spectral evolution.

In order to shed light on the issue, we have performed temperature dependent photoemission on a well known prototypical Kondo system, CeAl$_2$ (Kondo temperature T$_K$~5 K)[17]. To extract the Ce4$f$ contribution to the photoemission spectra, we have used resonant photoemission at the Ce 4$d$-4$f$ absorption threshold and subtracted the off-resonant (hν=112 eV) spectra from the on-resonant ones (hν=122 eV) (see supplementary information section A). The resulting angle integrated spectrum at 40 K is shown in Figure 1 along with a 0 K SIAM Ce4$f$ simulation[19], using the density of valence states for CeAl$_2$ calculated using the local density approximation (LDA) of the density functional theory. The purpose of this SIAM simulation is not to fit the experimental spectrum and extract the parameters of the Anderson model, but to get a qualitative understanding of the origin of individual features of the experimental spectrum. For this purpose, 0 K SIAM simulation gives good predictions since it describes the hybridization between the localized 4$f$ states and valence states. In the angle-integrated experimental spectrum, apart from the features corresponding to 4$f^0$, 4$f^1_{7/2}$ and 4$f^1_{5/2}$ final states which are well known in the literature[15,18], we observe weak features 'A' and 'B' which are well reproduced by the SIAM simulation taking into account the LDA valence band (VB) density of states for CeAl$_2$. The energy positions of features 'A' and 'B' correspond to the energy positions of strong peaks in the Ce4$f$ partial density of valence states for CeAl$_2$, which hybridize with the localized Ce4$f$ level[19]. Particularly, the features 'A', 'B', 4$f^1_{7/2}$ and 4$f^1_{5/2}$ correspond to the photoemission final state of the form 4$f^1$(VB)$^{-1}$, in which an electron from the valence band hops into the Ce4$f$ level filling up the hole produced by the 4$f$ photoemission process[20]. All features of the experimental spectrum are explained by the SIAM simulations qualitatively.

In Figure 2(a), we see that not only 4$f^1_{5/2}$ Fermi level peak but all the 4$f^1$(VB)$^{-1}$ final state features (including 'A' and 'B'), increase in intensity with respect to the 4$f^0$ feature, as the temperature is lowered. This convinces us that the origin of the temperature dependence of all the features 'A', 'B', $4f^1_{7/2}$ and $4f^1_{5/2}$ is the same, justifying the interpretation of features 'A' and 'B' as well screened features of $4f^1(VB)^{-1}$ final state configuration (see supplementary information section B). Reduction of the thermal broadening as discussed as a possible reason for a seeming increase of the Fermi level peak is only of the order of $k_BT$ and does not allow, therefore to explain the intensity variations of the much broader features 'A' and 'B'. On the other hand, an intrinsic increase of the hybridization strength as might be expected, e.g. from an increasing overlap of atomic orbitals in the context of thermal contraction of the crystal lattice, also does not explain the data. We have tried to simulate such an effect by increasing the hybridization parameter, $\Delta$, in our 0 K SIAM simulations. The results concerning the relative intensities of the individual spectral features with respect to the $4f^0$ feature are shown in the inset of Figure 1. As is obvious from this figure, increasing $\Delta$ affects particularly the intensity of the Fermi level peak. Moderate increase is also observed for the $4f^1_{7/2}$ sideband and peak 'B', while feature 'A' reveals a small decrease of intensity. Increasing $\Delta$, on the other hand, leads to a shift of the $4f^0$ component to higher binding energies which is experimentally not observed. It is thus clear that the observed changes in the experimental spectra are not because of the changes in the value of $\Delta$ due to thermal contraction of the lattice but represent the effects due to the many-body physics of the system. This is corroborated by the fact that the bulk physical properties of CeAl$_2$ display signatures of many-body physics within the temperature range of our investigation[21,22,23].

    The surface and bulk electronic structures in Kondo systems are known to differ in the strength of hybridization between 4$f$ and valence electrons which is larger in the bulk than at the surface due to higher atomic coordination in the crystal[24]. Although photoemission spectroscopy is basically a surface sensitive technique, it was demonstrated in ref. 25 that the differences between surface and bulk photoemission spectra are not significant in the case of low $T_K$ systems. Since this is a low $T_K$ system, we expect the results shown here to represent the bulk properties as well. Similar Ce4$f$

temperature evolution was observed in CeB$_6$, employing AlKα radiation (hν = 1486.6 eV) which leads to more bulk sensitive results[16]. This has also been observed in the bulk sensitive Ce 3$d$-4$f$ resonant photoemission results on CeSi$_2$ for the 4$f^1_{5/2}$ and 4$f^1_{7/2}$ features[26]. However, due to the broadness of the feature at 2 eV binding energy (4$f^0$ final state), we believe that the temperature evolution of the features in the intermediate range (between 4$f^1_{7/2}$ and 4$f^0$ feature) is not revealed. Hence the results shown in Figure 2 indicate genuine bulk Ce4$f$ spectral evolution.

Before going further, we would like to briefly discuss the possibility of the existence of singlet states in CeAl$_2$ above T$_K$. To address this issue, we explain the microscopic Kondo physics in a different way. The essence of Kondo physics is the increase of strength of the Kondo coupling between localized 4$f$ electrons and itinerant valence electrons at low temperatures. In the weak coupling regime (T > T$_K$), the Kondo coupling is small, hence the antiferromagnetic interaction between 4$f$ and valence electrons is weak. In this regime, a free valence electron comes, interacts antiferromagnetically with the 4$f$ electron for a short effective interaction time 'τ', and then gets scattered away as a free valence electron causing a spin flip in this process. In the strong coupling regime (T < T$_K$), the Kondo coupling is large, the valence electron cannot get scattered away because of the strong coupling which keeps the valence electron antiferromagnetically coupled to the 4$f$ electron forming a Kondo singlet. In this regime the effective interaction time 'τ' is very large and becomes infinite at 0 K. The infinite value of 'τ' amounts to stable Kondo singlet state which happens only at 0 K. At higher temperatures when 'τ' is finite, the singlet state is quasi-stable. So the temperature variation of the Kondo coupling gets reflected in the temperature variation of 'τ'. The two electrons' antiferromagnetic interaction (within the time 'τ'), both above and below T$_K$, is described by the same entangled singlet wave function. Therefore, even above T$_K$, entangled singlet states exist for a short interaction time 'τ'. 'τ' is of the order of 10$^{-13}$sec. for T$_K$~5 K and photoemission occurs on a much faster time scale (<10$^{-15}$ sec.). Assuming a logarithmic variation (from Figure 2(c)) of 'τ' with temperature, 'τ' reduces to about 10$^{-14}$sec. at 180 K. Since photoemission time scales (<$10^{-15}$ sec.) are still smaller than 'τ' even at 180 K, photoemission will sense these singlet correlations (within 'τ') and the measured photoemission spectrum will show features due to these entangled singlets. Here, we refer the reader to the existing huge literature on the Ce4$f$ photoemission spectra for different Kondo systems, many of them taken above and below $T_K$ for the respective systems[9,15,27], in which one observes that the Ce4$f$ photoemission spectral evolution from temperatures above $T_K$ to below $T_K$ is gradual without any sudden build up of the spectral intensity at the Fermi energy (Kondo resonance) below $T_K$; a fact well acknowledged by the photoemission community. It is well known that at temperatures above $T_K$, the Ce4$f$ photoemission spectrum shows a high temperature trail of the Kondo resonance (expected to occur only below $T_K$). All these results are consistent with our picture that there are singlet correlations (for a time 'τ') in Kondo systems even above $T_K$ and the photoemission technique senses them due to its faster dynamics.

The description of features 'A' and 'B' as 4$f^1$(VB)$^{-1}$ final state features, in the framework of multiorbital SIAM, leads to corresponding Kondo temperature scales for the features 'A' and 'B'[5]. In the same framework, it is known that the Kondo temperature scale increases as the 4$f^1$(VB)$^{-1}$ final state energy position moves towards higher binding energies. Since the temperature dependence of the 4$f^1$(VB)$^{-1}$ photoemission features is inversely proportional to their respective Kondo temperatures, this implies that the predicted temperature dependence becomes weaker and weaker as one goes towards higher binding energies. However from Figure 2(b) we see that all the features in the 4$f^1$(VB)$^{-1}$ final state configuration have a uniform temperature dependence. This is in direct contrast to the predictions of the multiorbital SIAM and clearly indicates that none of the features in the 4$f^1$(VB)$^{-1}$ final state configuration can be interpreted as Kondo resonance or its sidebands. Although we maintain that the features 'A', 'B', 4$f^1_{7/2}$ and 4$f^1_{5/2}$ arise due to screening of the 4$f$ photohole, we argue that the features cannot be interpreted as Kondo resonance or its sidebands. The reason that this simple experimental observation was missed by the photoemission community so far is due to the fact that only few photoemission reports exist (to the best of our knowledge) which try to study the temperature dependence of the whole Ce4$f$ emission (which includes both 4$f^0$ and 4$f^1$ final state features). Instead, almost all previous reports focus on features close to the Fermi energy.

CeAl$_2$ is a low T$_K$ system with localized Ce4$f$ magnetic moments[28]. Consequently, CeAl$_2$ lies at the Kondo limit of SIAM where the hybridization between Ce4$f$ and the valence electrons is small and thus the charge fluctuation effects are weak. In the Kondo limit of SIAM, the Anderson model can be transformed into the Kondo model which is described in terms of the Kondo coupling[29]. Therefore the electronic structure evolution of CeAl$_2$ with temperature can be described directly in the language of temperature variation of the Kondo coupling instead in the language of charge fluctuations as done in SIAM. Thus we have made an attempt to explain the increase of 4$f^1$(VB)$^{-1}$ final state features using the picture of temperature variation of the coupling between localized magnetic moments and valence electrons. We argue that the increase of features in 4$f^1$(VB)$^{-1}$ final state configuration represents increase of the coupling with reducing temperature. The reason for the intensity of 4$f^1$(VB)$^{-1}$ final state features to follow the increase of the coupling is due to the nature of the final states 4$f^1$(VB)$^{-1}$ which involve an electron transfer from VB to the 4$f$ level to screen the 4$f$ photohole. This transfer is sensitive to the strength of the coupling. The stronger the coupling between the 4$f$ and valence electrons, more will be the probability of screening the 4$f$ photohole by the valence electrons and larger will be the intensity of the 4$f^1$(VB)$^{-1}$ final state feature.

The photoemission spectrum obtained from a many-body system is understood to be the energy spectrum of the final states of the photohole created after the act of photoemission. Different final states of the photohole arise due to the possibility of different screening mechanisms in the system. The different final states are the projections of the photohole state on the eigenbasis of the system. Among one of the final states is the Kondo singlet state (represented by $|\uparrow\rangle_f|\downarrow\rangle_v-|\downarrow\rangle_f|\uparrow\rangle_v$ for a two electron system, where ↑, ↓ shows up and down spins respectively and *f,v* denotes 4*f* and valence electron respectively), giving rise to the Kondo resonance in electronic spectral function, which results from the spin flip scattering processes between 4*f* and valence electrons. Therefore one expects to observe the Kondo resonance in the photoemission spectrum. On the other hand, from Figure 2(a), we observe that there is no apparent signature of a sharp rise of the spectral intensity close to Fermi level as we reduce the temperature. We propose the following picture to explain the photoemission spectrum obtained from a Ce based Kondo system. The photoelectron detector extracts single particle information, e.g. kinetic energy etc., from the 4*f* photoelectron. Therefore the 4*f* photoelectron must be described in a well defined spin state (either only ↑ or ↓ spin but not a superposition of both ↑ and ↓ spins as in the Kondo singlet state). Therefore the corresponding 4*f* photohole will also have a well defined 4*f* spin state. Such a 4*f* photohole can have only those final states whose 4*f* spin state is the same as the spin state of the 4*f* photohole. Consequently any final state having a superposition of both ↑ and ↓ 4*f* spins (e.g. Kondo singlet state) cannot be a final state for such a 4*f* photohole. This concept amounts to the collapse[7] of the Kondo singlet wave function upon the act of photoelectron kinetic energy measurement. Based on this, we propose that the Kondo resonance cannot be seen in photoemission. In the collapsed state, the temperature dependence of all the features in the $4f^1(VB)^{-1}$ final state configuration will be due to the temperature dependence of the coupling between 4*f* electrons and valence electrons, which is similar to what we observe in Figure 2. On the contrary, if there was no collapse then one would get a sharp Kondo resonance at the Fermi level and therefore the Fermi level peak would show prominent temperature dependence as compared to other features of the $4f^1(VB)^{-1}$ final state configuration, as predicted by the spectral function[6]. A much more critical test for the proposal of wave function collapse would be to study the evolution of the Fermi surface with temperature, through photoemission spectroscopy. The formation of Kondo resonance would result in the expansion of the Fermi surface due to the contributions from 4*f* electrons to the Fermi volume, at low temperatures. However, to the best of our knowledge, no literature is available which reports the expansion of the Fermi surface in 4f based Kondo systems from high temperatures to low temperatures as observed through photoemission. We request the attention of the scientific community towards the proposal of the wave function collapse upon photoelectron kinetic energy measurement and invite discussions on it.

Additionally, while calculating the photoemission spectrum for any correlated electron system, the concept of wave function collapse must be incorporated. Since the wave function collapse is a classical phenomenon, the collapsed state should not be considered as a quantum state in order to calculate the photoemission spectrum (for a more elaborate discussion and better illustration, see supplementary information section C. Additionally, the comparison between a tunnelling spectrum and a photoemission spectrum as well as the observation of the quasiparticle spectral function in a superconductor have been discussed in supplementary information section C). We suggest that the calculation of the photoemission spectrum should be done semiclassically and not within fully quantum mechanical formalism.

Lastly, we are also aware of the current opinion that the transport based properties for $CeAl_2$ are better explained by a spin polaron based approach than by the Kondo model[21,22]. However, even the spin polaron resonance is a result of the spin flip scattering processes between 4f and valence electrons[30]. Since we argue that the final states corresponding to spin flip processes between 4f and valence electrons cannot be produced by the screening mechanism for the 4f photohole, hence even the spin polaron resonance cannot be observed in photoemission. Instead, the temperature dependence of the Ce4f spectrum is explained simply by the temperature dependence of the coupling between 4f and valence electrons which results from the hybridization between 4f and valence electrons. Such hybridization happens to be same in both the cases. Therefore we feel that it is difficult to resolve between Kondo or spin polaron based picture through photoemission. Thus the discussion presented in this manuscript holds in either case.

**METHODS SUMMARY**

Single crystals of $CeAl_2$ were prepared by the optical floating zone method and well characterized by X-ray diffraction and scanning electron microscopy in combination with energy dispersive x-ray analysis. For photoemission, the sample surfaces were prepared by fracturing the sample normal to the (001) direction in ultra high vacuum with a base pressure of $1\times10^{-10}$ mbar. The photoemission measurements were performed with the $1^2$ ARPES beamline at BESSY II synchrotron radiation facility in Berlin, Germany. Resonant photoemission was applied exploiting the giant increase in the 4$f$ photoemission cross-section close the Ce4$d$-4$f$ absorption edge at photon energy of 122 eV to study selectively the Ce4$f$ states. Valence band spectra were measured at off-resonance using photon energy of 112 eV. At every temperature, a fresh surface was prepared in order to obtain clean spectra. The spectra were collected within about 15 minutes after the surface preparation, to minimize the contamination of the reactive surfaces of Ce compounds. The lattice parameter for this cubic Laves phase compound is $a = 8.06$ Å which leads to the Brillouin zone length of about 16.3° in terms of the electron emission angle, at the photoelectron kinetic energy of 118 eV. The angle integrated spectra presented in this manuscript were obtained by integrating the photoemission intensity within ±15° of the electron emission angle, much larger than the Brillouin zone length. Together with the fact that fracturing the sample generates rough surfaces, this leads to integration over a large effective k-space of the sample at the photon energies used for the measurements.

**Supplementary Information** is linked to the online version of the paper at www.nature.com/nature.

**Acknowledgements** We thank Prof. C. Laubschat, Prof. N.E. Sluchanko and Prof. O. Gunnarsson for fruitful discussions. We thank Prof. B. Büchner, Dr. S. Wurmehl for sample preparation and characterization. We thank Prof. Y. Kucherenko for doing the SIAM simulations. S.P. acknowledges financial assistance from Alexander von Humboldt Foundation, Germany. We thank C.G.F. Blum for technical support. We thank the members of the mechanical workshop of Institut für Festkörperphysik, TU Dresden, Germany, for their great help in preparing specialized sample holder for the photoemission measurements.

**Author Contributions** S.P. planned the experiments. S.P and A.G. conducted the experiments. A.O. prepared samples and characterized them. S.P. did the data analysis and manuscript preparation.

**Competing Interests** The authors declare that they have no competing financial interests.

**Correspondence** Correspondence and requests for materials should be addressed to S.P. (swapnil@physik.phy.tu-dresden.de)

**Figure 1 | Comparison between 0 K SIAM simulation and experimental 40 K Ce4$f$ spectrum.** The experimental spectrum shows 5 final state features, $4f^0$, 'A', 'B', $4f^1_{7/2}$ and $4f^1_{5/2}$. The feature $4f^0$ is the poorly screened photoemission feature while the features $4f^1_{7/2}$ and $4f^1_{5/2}$ are the well screened photoemission features (split by the spin orbit interaction) and are well known in the literature[15,18]. The features $4f^0$, 'A' and 'B' are well reproduced by the SIAM simulation[19] taking into account the density of valence states for CeAl$_2$ calculated using the local density approximation (LDA) of the density functional theory. The feature $4f^1_{7/2}$ is somewhat underestimated in the simulations. The Fermi energy feature $4f^1_{5/2}$ is not properly simulated by the 0 K simulation due to the fact that the thermal broadening of features influence dominantly the experimental spectra close to the Fermi energy. SIAM parameters used for calculations: $\varepsilon_f$ (energy of the bare 4$f$ level) = -1.85 eV, $U_{ff}$ (Coulomb repulsion between 4$f$ electrons) = 6 eV, Spin-

orbit splitting for Cerium = 0.28 eV, $\Delta$ (overlap between 4$f$ and valence electrons) = 0.95 eV. The inset (*) shows the relative intensity of the final state features 'A', 'B', $4f^1_{7/2}$ and $4f^1_{5/2}$ with respect to the intensity of the $4f^0$ feature, estimated within the SIAM simulations with values of $\Delta$ within nearly ±15% about 0.95 eV (value of $\Delta$ used for comparison with the experimental spectrum. The value of $\varepsilon_f$ was kept fixed to -1.85 eV which is the value of $\varepsilon_f$ used for the comparison with the experimental spectrum). We see that the behaviour of feature 'A' with the increase in $\Delta$, is different as compared to the behaviour shown by other features. Relative intensity for 'A' weakly decreases with the increase of $\Delta$ whereas the relative intensity for all other features seem to increase.

**Figure 2 | Temperature evolution of Ce4$f$ spectra.** (a) The spectra for all temperatures have been normalized at the peak intensity of the $4f^0$ feature to study the changes in the intensity of the $4f^1(VB)^{-1}$ final state features with temperature. All the features in the $4f^1(VB)^{-1}$ final state configuration ('A', 'B', $4f^1_{7/2}$ and $4f^1_{5/2}$) show increase at low temperatures. In the inset, the near Fermi energy region is expanded and compared with the Fermi distribution function at 40 K and 180 K to show that the conventional thermal broadening does not account for the decrease of the intensity of the $4f^1(VB)^{-1}$ final state features with increasing temperature. (b) The spectra normalized to the intensity at 0.5 eV binding energy reveal a uniform temperature dependence for all the features in the $4f^1(VB)^{-1}$ final state configuration ('A', 'B', $4f^1_{7/2}$ and $4f^1_{5/2}$). The inset shows the expanded region close to Fermi energy. (c) Increase of the $4f^1(VB)^{-1}$ final state features at low temperatures is estimated by integrating the area under the difference curve, Ce4$f$(T)-Ce4$f$(180 K) from Figure 2(a), between 0.3 eV above Fermi energy and 2 eV binding energy. The intensity increase of the $4f^1(VB)^{-1}$ final state features, which is due to the increase of the coupling between 4$f$ and valence electrons at low temperatures, follows an approximately logarithmic behaviour as a function of temperature as shown by the red dashed line.

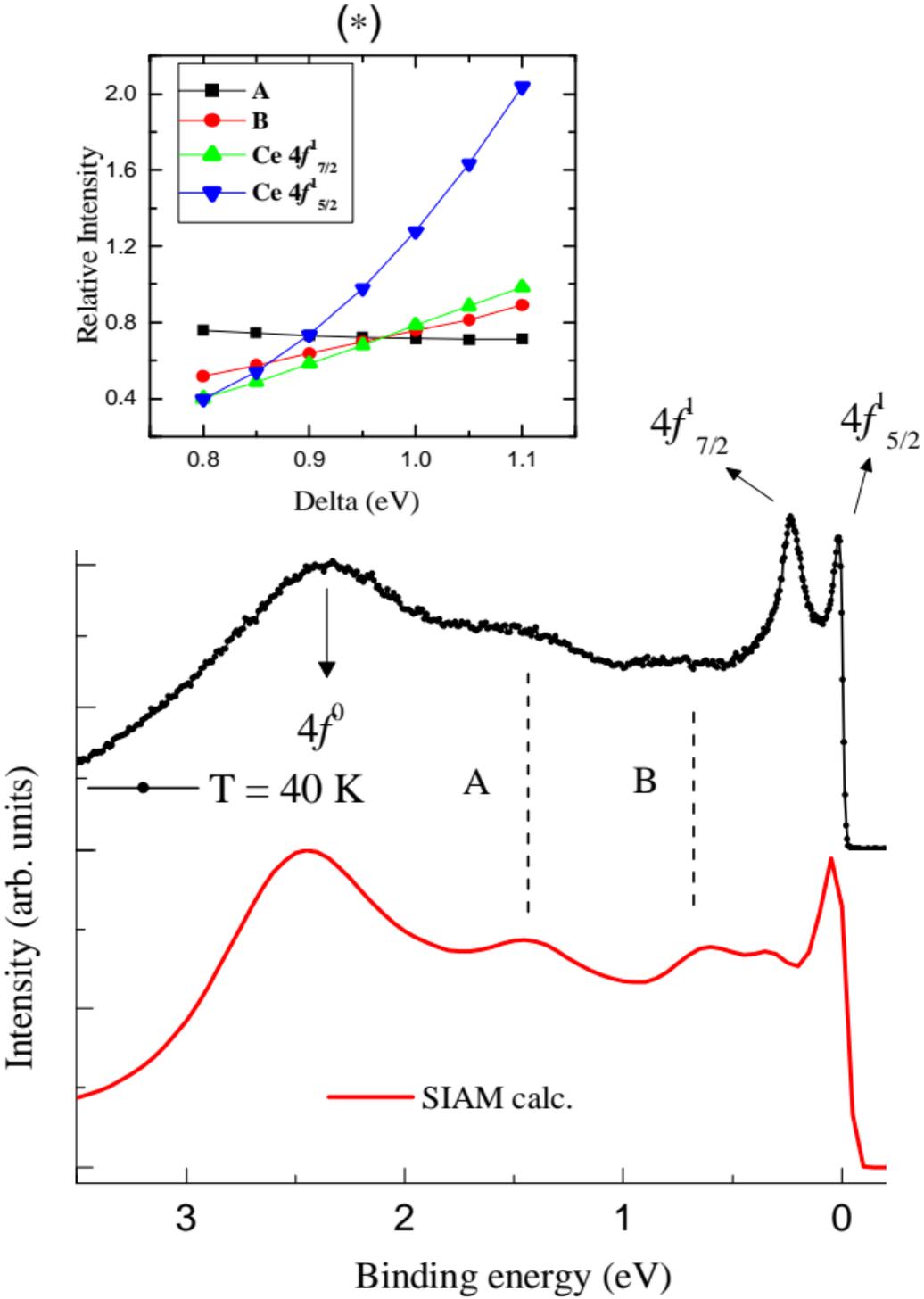

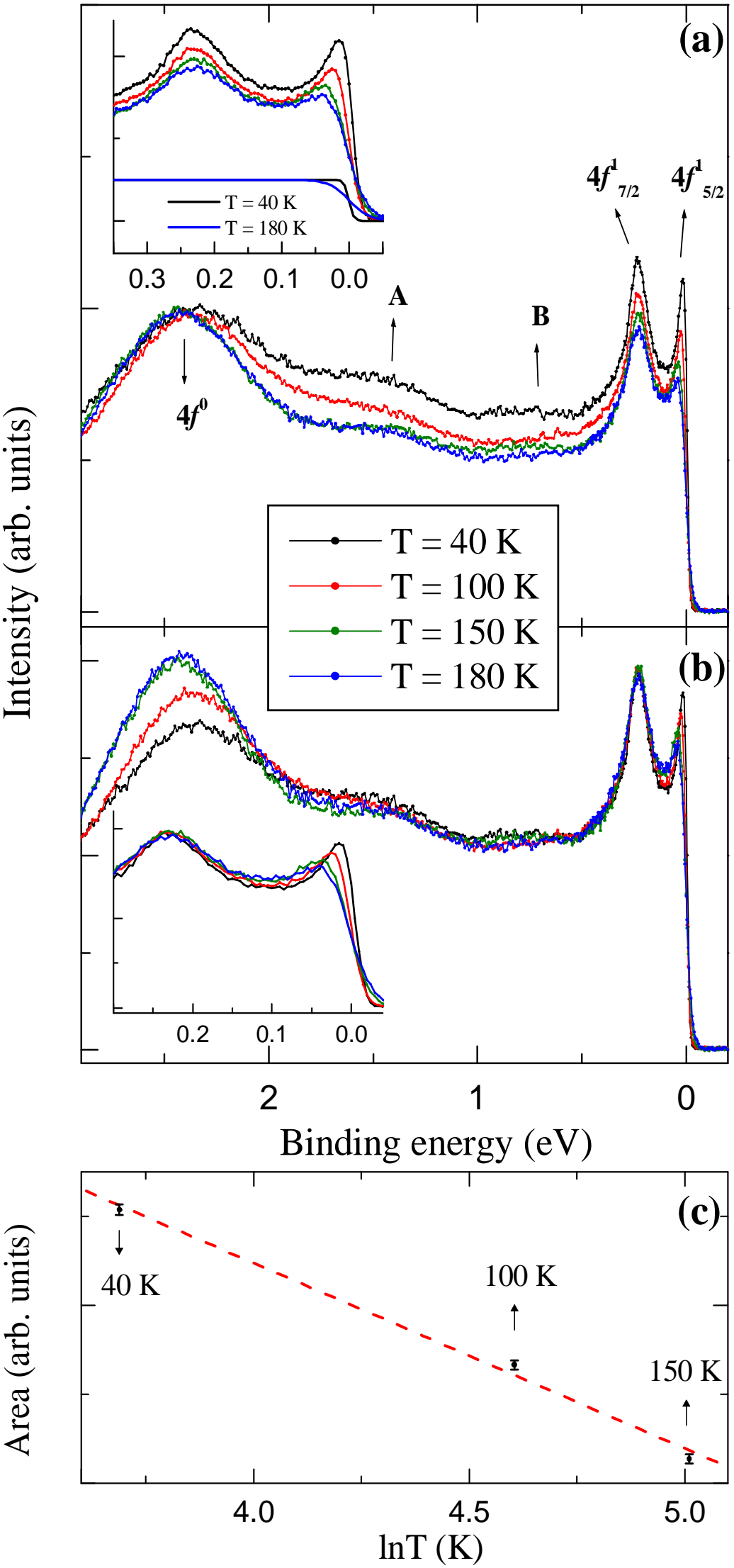

# Supplementary Information

# A] Extraction of the Ce 4*f* spectrum.

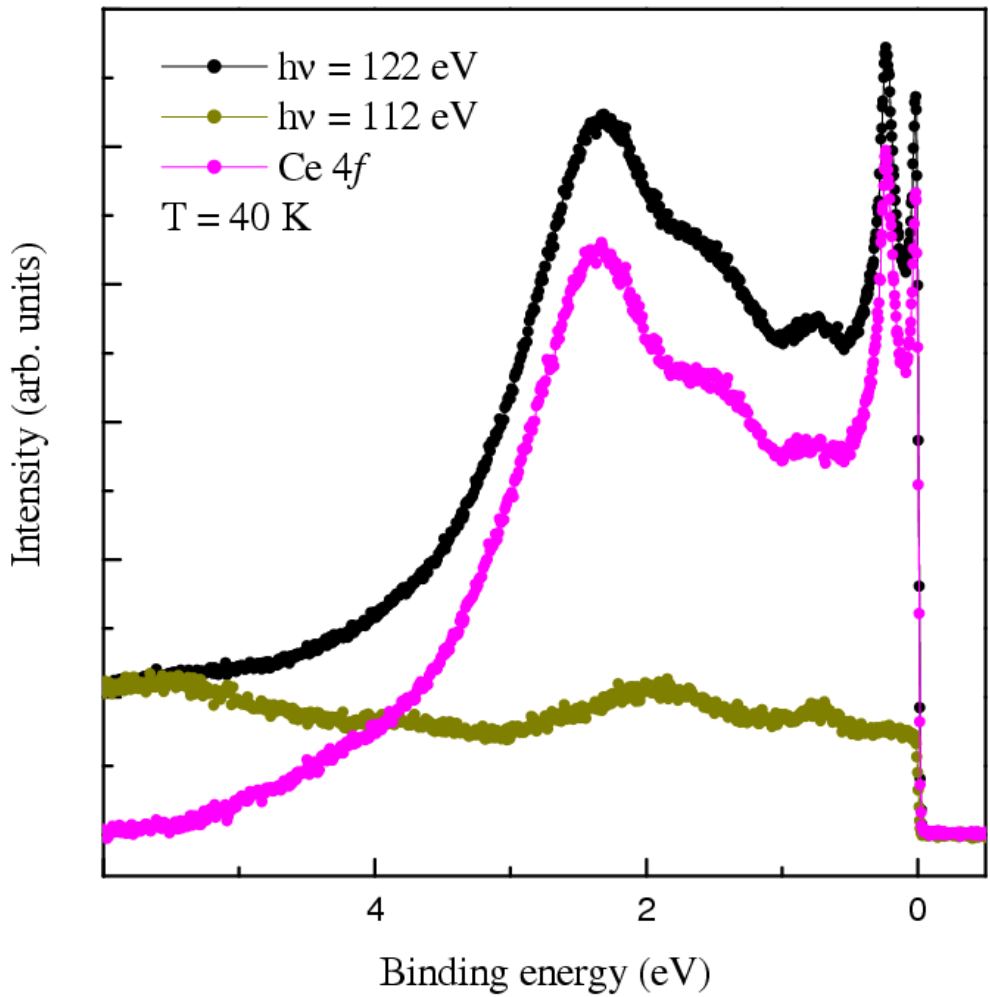

**Figure S1:** On-resonant (hν = 122 eV) photoemission spectrum plotted with off-resonant (hν = 112 eV) photoemission spectrum. Both are normalized at 6 eV binding energy. The extracted Ce4*f* spectrum, obtained by subtracting the off-resonant spectrum from the on-resonant spectrum, is shown.

# B] Comment on the expanse of the screened 4*f* photoemission features.

Features 'A' and 'B' are generally interpreted as a part of the ionization peak ($4f^0$ final state hybridized with the *spd* valence band of $CeAl_2$). However the different relative temperature evolution of the features 'A' and 'B' with respect to the highest binding energy feature $4f^0$ (see Figure 2(a)) compels us to ascribe different photoemission origins to the feature $4f^0$ from features 'A' and 'B'. Since the temperature evolution of features 'A' and 'B' follows the temperature evolution of $4f^1_{7/2}$ and $4f^1_{5/2}$, we ascribe them to $4f^1(VB)^{-1}$ final state configuration. So the $4f^1(VB)^{-1}$ final states extend from the Fermi energy to the highest binding energy feature ($4f^0$ final state). This is in contrast with the current understanding of 4*f* photoemission features ascribing only the features $4f^1_{7/2}$ and $4f^1_{5/2}$ to $4f^1(VB)^{-1}$ final state configuration. The features 'A' and 'B' are broader due to significant 4*f* hybridization with the broad *spd* valence band whereas the final states $4f^1_{7/2}$ and $4f^1_{5/2}$ have mostly atomic character and hence appear as narrow peaks.

The spectral decomposition ($4f^0$ or $4f^1(VB)^{-1}$ final state) based on theoretical models depends largely on the type of model and the basis sets used for the calculations. However, it is generally true that the $4f^0$ final state spectrum gives a maximum contribution to the highest binding energy feature in the Ce4*f* photoemission spectrum and it gradually diminishes as one goes towards the Fermi energy. Whereas the $4f^1(VB)^{-1}$ final state spectrum gives a maximum contribution to the Fermi level peak and it gradually diminishes as one goes towards higher binding energies. Therefore the 4*f* photoemission features between the Fermi level and highest binding energy feature seem to have a mixed character of the final state ($4f^0$ and $4f^1(VB)^{-1}$). While the $4f^0$ final state is inherently temperature independent, the $4f^1(VB)^{-1}$ final state shows a temperature dependence characteristic of the Anderson (or any other theoretical model) model.

# C] More on the wave function collapse upon photoelectron kinetic energy measurement

In the theory of quantum mechanics, the phenomenon of ascribing a well defined spin state to the photoelectron can be described as the collapse of the entangled Kondo singlet wave function upon the act of kinetic energy measurement. For simplicity, let us assume that the Kondo singlet is formed between a 4$f$ electron and one valence electron. Then the singlet wave function looks like $|\uparrow\rangle_f|\downarrow\rangle_v - |\downarrow\rangle_f|\uparrow\rangle_v$, where $\uparrow, \downarrow$ shows up and down spin respectively and $f,v$ denotes 4$f$ and valence electron respectively. The following discussion explains the concept of the wave function collapse for a two electron system. The act of photoelectron kinetic energy measurement in the process of Ce4$f$ resonant photoemission spectroscopy, determines the energy properties of one of the components of the electron pair forming the Kondo singlet (the other component is left inside the sample). By the word component, we refer to either the 4$f$ or valence electron. By the act of measuring the kinetic energy of the photoelectron, we also indirectly determine (but not detect; we can only detect the spin of the photoelectron if we do a spin resolved experiment) the spin of the photoelectron, since the measured photoelectron has a well defined spin state and not a superposed spin state as in the Kondo singlet state. When the measurement is performed on one component of the pair, the wave function collapses and is no more entangled. The collapsed wave function will be described as $|\uparrow\rangle_f|\downarrow\rangle_v$ (or $|\downarrow\rangle_f|\uparrow\rangle_v$ depending on the spin $\uparrow$ or $\downarrow$ of the 4$f$ photoelectron respectively). Since this collapsed wave function is no more an entangled fermionic singlet wave function, Kondo resonance is not observed in photoemission. This is exactly the same picture as proposed for explaining the measurements done on an entangled EPR (Einstein, Podolsky, Rosen) pair. E.g. if we do a spin measurement of one of the components of the electron pair forming an entangled spin singlet state then the spin of the other component is automatically determined. In other words, the wave function collapses to the one which gives the measured spin values to both the components and the collapsed wave function is not entangled.

{Note: The wave function collapse generated by the act of measurement makes the photoelectron, an indistinguishable fermionic particle, appear to the experimenter as a distinguishable particle! The act of kinetic energy measurement distinguishes the photoelectron from other indistinguishable electrons. The indistinguishability between two particles can be understood by examining the properties of the two particles' quantum state upon particle exchange. In the case of the Kondo singlet state $|\uparrow\rangle_f|\downarrow\rangle_v - |\downarrow\rangle_f|\uparrow\rangle_v$, particle exchange does not change the wave function except for an overall phase factor. Therefore we say that the two particles are indistinguishable. However, in the collapsed state the particle exchange gives rise to a completely new state. Let us suppose that we measured the 4$f$ electron in $\uparrow$ spin. Then the collapsed wave function becomes $|\uparrow\rangle_f|\downarrow\rangle_v$. Upon particle exchange we get a new state $|\downarrow\rangle_f|\uparrow\rangle_v$ in which the 4$f$ electron is in $\downarrow$ spin. This is not what we observed in our experiment and hence this state has become distinct from $|\uparrow\rangle_f|\downarrow\rangle_v$. This means that the particles have become distinguishable. Thus the phenomenon of wave function collapse following the measurement of a single particle property, introduces distinguishability amongst quantum particles! This has not been explicitly elaborated in the previous literature but this is the physical meaning contained in the phenomenon of wave function collapse. However, the collapse is only an apparent phenomenon "at the moment" of measurement of single particle properties. After the measurement is performed, when the electron passes through the microchannel plate towards earth it again behaves as an indistinguishable electron with respect to other electrons of the microchannel plate.}

**Comment on the calculation of the photoemission spectrum considering wave function collapse:**

The different quantum states for a two electron system are a singlet state and 3 triplet states. The singlet and triplet states constitute the basis for the two electron Hilbert space. The singlet and triplet states are:

Singlet state '$S_{sin}$' = $\dfrac{|\uparrow\rangle_f|\downarrow\rangle_v - |\downarrow\rangle_f|\uparrow\rangle_v}{\sqrt{2}}$

Triplet states '$S_{tri}$' = $\begin{cases} S_{\uparrow\uparrow} = |\uparrow\rangle_f|\uparrow\rangle_v \\ S_{\uparrow\downarrow} = \dfrac{|\uparrow\rangle_f|\downarrow\rangle_v + |\downarrow\rangle_f|\uparrow\rangle_v}{\sqrt{2}} \\ S_{\downarrow\downarrow} = |\downarrow\rangle_f|\downarrow\rangle_v \end{cases}$

Let us consider that a Kondo singlet, $|\uparrow\rangle_f|\downarrow\rangle_v - |\downarrow\rangle_f|\uparrow\rangle_v$, is formed between one 4$f$ and one valence electron. Let us, for simplicity, neglect the internal structure of the Ce4$f$ manifold i.e. spin orbit and crystal field splittings of Ce4$f$ level. Consider 4$f$ photoemission from the Kondo singlet state. Upon photoemission the 4$f$ photohole can have poorly screened (4$f^0$) and well screened (4$f^1$) final states. In the case of 4$f^1$ final states, a valence electron hops into the 4$f$ photohole. If $|\psi\rangle$ denotes the well screened 4$f$ photohole state then it is currently believed that $|\psi\rangle$ can be written as,

$$|\psi\rangle = a\{S_{sin}\} + b\{S_{tri}\} = a\left(\dfrac{|\uparrow\rangle_f|\downarrow\rangle_v - |\downarrow\rangle_f|\uparrow\rangle_v}{\sqrt{2}}\right) + b\left\{\left(|\uparrow\rangle_f|\uparrow\rangle_v\right) + \left(\dfrac{|\uparrow\rangle_f|\downarrow\rangle_v + |\downarrow\rangle_f|\uparrow\rangle_v}{\sqrt{2}}\right) + \left(|\downarrow\rangle_f|\downarrow\rangle_v\right)\right\},$$

where $a$, $b$ are complex numbers. The above expression shows the two well screened 4$f$ features in the 4$f$ photoemission spectrum; the "nonmagnetic" 4$f^1$ final state (Kondo resonance) feature (from $S_{sin}$) and the "magnetic" 4$f^1$ final state features (from $S_{tri}$) both separated by the energy $\sim k_B T_K$. This is the current theoretical picture of photoemission without considering the phenomenon of wave function collapse upon photoelectron kinetic energy measurement.

<u>The theoretical picture considering the wave function collapse will look like</u>:

Suppose that after photoemission from the Kondo singlet state, the 4$f$ photoelectron is detected in ↑ spin (without loss of generality). Then the Kondo singlet wave function collapses to $|\uparrow\rangle_f|\downarrow\rangle_v$. The collapse leaves the 4$f$ photohole in ↑ spin. This ↑ spin 4$f$ photohole can only be screened by a valence electron with ↑ spin. After screening, the 4$f$ electron (occupying the 4$f$ photohole) also has a ↑ spin state, therefore it cannot form a singlet state (containing superposition of both ↑ and ↓ 4$f$ spins) or a triplet state (containing superposition of both ↑ and ↓ 4$f$ spins as in $S_{\uparrow\downarrow}$ and containing ↓ 4$f$ spin as in $S_{\downarrow\downarrow}$) with the valence electron! Therefore the well screened collapsed 4$f$ photohole state must be written as $|\psi_{collapse}\rangle = |\uparrow\rangle_f|\downarrow\rangle_v$ (but not $|\psi_{collapse}\rangle = |\uparrow\rangle_f|\downarrow\rangle_v - |\downarrow\rangle_f|\uparrow\rangle_v$ etc.). In that case;

$$|\psi_{collapse}\rangle = |\uparrow\rangle_f|\downarrow\rangle_v \neq a\left(\dfrac{|\uparrow\rangle_f|\downarrow\rangle_v - |\downarrow\rangle_f|\uparrow\rangle_v}{\sqrt{2}}\right) + b\left\{\left(|\uparrow\rangle_f|\uparrow\rangle_v\right) + \left(\dfrac{|\uparrow\rangle_f|\downarrow\rangle_v + |\downarrow\rangle_f|\uparrow\rangle_v}{\sqrt{2}}\right) + \left(|\downarrow\rangle_f|\downarrow\rangle_v\right)\right\}$$

The well screened collapsed 4*f* photohole state cannot be written as an expansion in terms of $S_{sin}$ and $S_{tri}$ for any nonzero values of *a* and *b*. $|\psi_{collapse}\rangle$ is not a quantum state (its wave function does not obey the symmetry properties required for the wave functions describing indistinguishable particles) and it cannot be expanded in terms of $S_{sin}$ and $S_{tri}$. Therefore the well screened final states for such a collapsed 4*f* photohole cannot be labelled as "singlet" or "triplet".

**Comparison between tunnelling spectrum and photoemission spectrum:**

The electron entering the photoelectron detector and whose single particle properties are measured must be described in a collapsed state whereas the same electron when contributes to the electric current measurement (tunnelling current in case of tunnelling spectroscopy) should not be described in a collapsed state as its entanglement with other electrons of the solid is preserved. Therefore tunnelling spectrum will show many-body physics in its entirety. Photoelectron detector distinguishes the photoelectron from other indistinguishable electrons in the solid, making the photoelectron appear to the experimenter as a distinguishable particle, whereas tunnelling spectroscopy maintains indistinguishability amongst all the electrons of the solid preserving their quantum behaviour from the point of view of the experimenter.

Indeed, Kondo resonances have been observed in tunnelling spectroscopy (see Phys. Rev. Lett **88**, 096804 (2002); Science **280**, 567 (1998); Science **289**, 2105-2108 (2000)).

Alternatively, there is no requirement of the spin dependent constraints (contrary to the case of the photohole whose spin has to be well defined) while calculating the many-body spectral function describing the tunnelling spectrum. Hence the tunnelling spectrum will show complete manifestations of the spin many-body physics of the system. This is the subtle difference between both the experimental methods, although traditionally both of them have been known to give the same information i.e. electronic spectral function of the solids.

Above discussion highlights the differences in the information obtained, between different types of measurements performed on the same quantum system. One type of measurement is the detection and study of the properties of individual electrons, and the other type of measurement is the study of a property of the whole many electron system without any specific enquiry into the properties of the individual electrons. One more example of the second type is specific heat measurements in which a heat pulse is applied to the solid and the resulting temperature is measured. But while measuring the temperature the experimenter has absolutely no information on the properties of the individual electrons. This experiment does not study individual electrons contrary to photoemission spectroscopy which studies individual electrons. More formally, the photoelectron must have a collapsed state upon kinetic energy measurement whereas the same electron will maintain its entanglement with other electrons of the solid during the course of specific heat measurements. Therefore, the specific heat measurements reveal complete manifestations of the entangled states in Kondo systems through the demonstration of heavy electrons.

We hypothesize that any experiment involving either electron in or electron out (and the study of single electron properties like kinetic energy etc.) will not be able to show signatures of many-body physics of the system. In the case of photoemission spectroscopy, this is straightforward due to the wave function collapse upon measurement of the photoelectron kinetic energy. However, in the case of inverse photoemission spectroscopy where we detect photons instead of electrons, one would be tempted to think that the wave function collapse will not occur and hence the many-body physics of the electrons can be studied. But here too one may not study the many-body physics since the electron coming out from the electron gun and entering the sample has a definite kinetic energy and hence a well defined spin state, therefore such an electron cannot occupy a state inside the solid sample having a superposition of both ↑ and ↓ spins (e.g. Kondo singlet state) and hence the many-body physics may not be studied.

**Why we see quasiparticle spectral function in a superconductor?**

Following the discussion of the wave function collapse upon photoelectron kinetic energy measurement and subsequent inability to see the many-body features in photoemission, a natural question arises as why do we see the quasiparticle spectral function of a BCS superconductor (along with a superconducting gap), in the photoemission spectrum when the electrons forming a Cooper pair are also in a singlet state. We try to explain this issue qualitatively, conceptually without the use of the theoretical/mathematical rigour.

Cooper pairs are formed between an electron pair due to the involvement of phonons, which can be described in terms of an effective single particle picture of weakly interacting Bogoliubov quasiparticles. Each quasiparticle contains one electron; hence the electron pair in the Cooper pair singlet state is effectively decoupled into two weakly interacting Bogoliubov quasiparticles. During the photoemission, when an electron from one quasiparticle is emitted and measured, the electron inside the other quasiparticle cannot "feel" the effect of the measurement of the first electron due to the weak interaction. Hence the photoemission spectrum from such a Cooper pair state does not show significant departures from the predictions of the spectral function calculations.

However, heavy quasiparticles (heavy fermions) are also formed in $4f$ based Kondo systems. Here the quasiparticle means "the $4f$ electron coupled to the cloud of the valence electrons surrounding it". In the heavy quasiparticle, the $4f$ electron is strongly interacting with the valence electrons (that is why we get heavy masses for valence electrons). In the Kondo singlet state arising from the $4f$ and valence electron, both the electrons influence the same quasiparticle, unlike the case of a Cooper pair where we get two different quasiparticles from an electron pair. Hence during photoemission, when a $4f$ electron is emitted and measured, the valence electron "feels" the effect of the measurement of the $4f$ electron due to the strong interaction! And hence the observed photoemission spectrum will show significant departures from the predictions of the many-body spectral function calculations.

# Appendix

# Continuum description for the localized electrons hybridizing with the valence band

Swapnil Patil*

*Department of Condensed Matter Physics and Materials Science, Tata Institute of Fundamental Research, Homi Bhabha Road, Colaba, Mumbai 400005, India*

*Email: swapnil.patil@tifr.res.in

**Abstract**

In this short communication, intended as the author's viewpoint, we propose a continuum description for the localized electrons hybridizing with the valence band. Normally their theoretical understanding is obtained within the framework of Anderson-based models like the single impurity Anderson model (SIAM) etc. SIAM is extensively used while calculating the local electron spectral function in which the localized impurity state hybridizing with the valence band via a hopping interaction is expanded within the atomic basis. We argue that under the influence of such hybridization the valence band 'spreads' the localized impurity state over its width upto the binding energy (BE) position of the 'ionization' peak in the local electron spectral function. As a result the expansion needs to be performed within a continuum basis and hence the many-body features would accordingly appear continuously across the aforementioned valence band energy range.

**Discussion**

Recently we have published a paper [1] highlighting that remarkably strong temperature dependence is visible in the Ce4$f$ photoemission spectra of a Ce-based Kondo system CeAl$_2$ beyond 0.5eV BE. SIAM [2], however, predicts temperature dependence for a Ce impurity to be limited within 0.5 eV BE [3-9]. This is because the maximum splitting of the ground state for a Ce4$f$ electron is of the order 0.28 eV which is due to the spin orbit (SO) splitting. Therefore the maximum value of BE over which the SIAM temperature dependence is predicted is 0.28eV. A lot of experimental evidence has been obtained in photoemission spectroscopy (PES) which has substantiated the energy splitting of the Ce4$f$ ground state generating a Ce4$f$ PES feature at ~0.28 eV BE corresponding to the SO split excited state along with Ce4$f$ PES features corresponding to the crystalline electric field (CEF) splitting of the ground state[6-16]. This fine structure of the Ce4$f$ PES feature reflects a typical atomic-like state for the Ce4$f$ electron. Therefore it was justified to use the atomic basis for expanding the Ce4$f$-impurity state within the framework of SIAM. All the excited states of the impurity are populated in final state of the photoemission since the photo-hole state allows a finite projection over the excited states of the impurity. Accordingly each final state generates a feature in the Ce4$f$ PES spectra giving rise to its fine structure. The PES features arising from the excited states appear at higher BE's and are known as Kondo sidebands while the PES feature arising from the ground state is called the Kondo resonance. Every such feature is indeed a many-body resonant excitation with its characteristic temperature scale (called Kondo sideband temperature or Kondo temperature respectively) associated with it [7, 17]. However experimental studies probing the temperature dependence of these Kondo sidebands have been rare. Instead only Kondo resonance has been extensively studied for its temperature dependence so far.

In our paper, we have specifically looked for such temperature dependence of Ce4$f$ PES features at higher BE's and have found a strong temperature dependence there which is very puzzling [1]. Surprisingly we find strong temperature dependence for states beyond the SO Kondo sideband which appear to come from the valence band density of states (DOS). However this anomalous temperature dependence arises within the Ce4$f$ spectral function itself. Thus the broad features beyond 0.5 eV BE (which are generally believed to be valence band features) indeed have a 4$f$ symmetry at the Ce-atomic site as they are obtained in the Ce4$f$ sensitive experimental probe [1]. Furthermore, they can be explained by a SIAM calculation (which calculates Ce4$f$-impurity spectral function) taking into account the LDA valence band of CeAl$_2$, where these features actually reflect the peaks in the Ce4$f$ partial density of valence states of CeAl$_2$ which hybridize with the localized Ce4$f$ level, thus supporting our claim that these anomalous Ce4$f$ features are indeed a part of the Ce4$f$ spectral function. Thus the Ce4$f$ temperature dependence can be seen to extend continuously from the Fermi level (E$_F$) till the BE position of the ionization peak (4$f^0$ final state feature) which is very surprising and inconsistent with the current SIAM predictions.

Our observation of the broad features beyond 0.5 eV BE and within 0.5 eV BE displaying similar temperature evolution *compels* us to treat the PES final states for both these features *on the same theoretical footing*. Thus we are tempted to question the validity of the atomic basis set (which explains the temperature dependence only upto the SO Kondo sideband) for the expansion of Ce4$f$-impurity states. In fact we tend to believe that a continuum description for the Ce4$f$-impurity state is required in order to explain our experimental results. Such a continuum basis appears to be naturally justified due to a finite hybridization between the Ce4$f$ electron and the valence band converting the atomic Ce4$f$ state into a 'hybrid' state continuously extending from E$_F$ upto the BE position of the ionization peak. In other words the hybridization between the atomic Ce4$f$ state and the valence band does not allow us to view the Ce4$f$ state as a distinct entity having its unique spectral profile different from the valence band spectral profile. In effect it no more remains an atomic Ce4$f$-impurity state but rather becomes a 'band' Ce4$f$-impurity state. So a 'continuous' DOS description for the Ce4$f$-impurity band is required instead of a discrete atomic multiplet description. In such a band scenario the existence of sharp peaks in the PES spectra corresponding to the atomic energy level splitting of the Ce4$f$ state can then be understood as peaks in DOS arising from it. With increasing hybridization these peaks gradually vanish into the continuum band (akin to a loss of atomic behavior and emergence of band behavior) which has been experimentally observed [18].

In the event of this continuum description for the hybridized Ce4$f$ state, the many-body character of the PES features would naturally extend continuously upto the BE position of the ionization peak. Thus the Kondo sidebands would also then form a continuum band extending upto the ionization peak. The Kondo sideband temperature scale monotonically increases with increasing BE as has been concluded from recent theoretical results which have clearly demonstrated (using a two level system with variable energy spacing between them) that the Kondo temperature scale for the excited state increases monotonically with its increasing energy separation from the ground state [17]. Extending this notion to the continuum case, we hypothetically pin the lower state of the two level system to E$_F$ while the upper state is kept variable in energy simulating various energy positions for individual states within the continuum and observe that the Kondo (sideband) temperature scale monotonically increases with increasing BE even for the continuum. Thus in this continuum Ce4*f* scenario the Kondo sideband temperature scale becomes a monotonically increasing 'continuous' function of BE as against the atomic Ce4*f* scenario for which the Kondo sideband temperature scale is a monotonically increasing 'discrete' function of BE.

**Summary**

In summary, we argue that a localized state hybridized with the valence band should not be expanded in the atomic basis. Instead a continuum basis is required for calculating its spectral properties. This is because the hybridization does not allow an existence of a unique spectral profile for the localized electron which is distinct from the spectral profile for the valence band. In fact the spectral profile for the localized electron merges with that of the valence band. Effectively this would lead to a continuum distribution for the corresponding many-body features. E.g. in Kondo systems this would lead to the Kondo sideband temperature scale increasing monotonically in a continuous manner with increasing BE in its local electron spectral function.